# Effects of Daily News Sentiment on Stock Price Forecasting

R. Satish Srinivas
Data Science
Tiger Analytics
Chenai, India
rsatish57@gmail.com

Rakesh Babu Gedela
Data Science
Genpact
Hyderabad, India
rakesh.smiley@gmail.com

Rohit Sabu
Data Science
PWC
Bangalore, India
rohitsabu96@gmail.com

Avik Das
Data Science
Techno India University
Kolkatta, India
musicvik@gmail.com

Gourab Nath
Department of Data Science
Praxis Business School
Bangalore, India
gourab@praxis.ac.in

Vaibhav Datla
Data Science
PWC
Bangalore, India
vaibhavvarmadv@gmail.com

*Abstract*— **Predicting future prices of a stock is an arduous task to perform. However, incorporating additional elements can significantly improve our predictions, rather than relying solely on a stock's historical price data to forecast its future price. Studies have demonstrated that investor sentiment, which is impacted by daily news about the company, can have a significant impact on stock price swings. There are numerous sources from which we can get this information, but they are cluttered with a lot of noise, making it difficult to accurately extract the sentiments from them. Hence the focus of our research is to design an efficient system to capture the sentiments from the news about the NITY50 stocks and investigate how much the financial news sentiment of these stocks are affecting their prices over a period of time. This paper presents a robust data collection and preprocessing framework to create a news database for a timeline of around 3.7 years, consisting of almost half a million news articles. We also capture the stock price information for this timeline and create multiple time series data, that include the sentiment scores from various sections of the article, calculated using different sentiment libraries. Based on this, we fit several LSTM models to forecast the stock prices, with and without using the sentiment scores as features and compare their performances.**

## I. INTRODUCTION

Modern financial theories assume investors are homogeneous and rational, which is erroneous. Behavioural finance, in contrast, explores investor behaviour in the stock market and its possible effects on stock price movements. According to [1], investors are susceptible to sentiment. Based on the assumption that stock price movements can be influenced by investor sentiment, the current study seeks to investigate how financial news-based sentiment can improve stock price prediction in the Indian market.

Most players in the financial market, including investors, are sentimental and emotionally driven. Financial judgments based on emotion are inherently flawed and at a contradiction with the objectives of investments, according to behavioural finance. During the 2008 and 2020 financial collapse, investors exhibited extreme herding in the financial market. It is a highly developed and established fact that the investor sentiment and the stock market movements are related [2].

Our research explores the methods of sentiment analysis which is primarily focusing on popular financial dictionaries and forecast the stock prices using an LSTM model. The primary objective of this work is to generate one or more time series that capture the daily news sentiment scores generated based on a daily national business newspaper and use such series to design an efficient forecasting model to forecast stock prices. We believe that the most appropriate choice for a national newspaper is Economics Times (ET). ET is India's largest business daily that publishes more than 300 news articles daily. One of the primary steps of this work includes designing a robust data collection framework and cleaning technique to collect the daily news data of the last three odd years to create a news database of almost half a million news articles. We record the date, category, article heading, article synopsis and article full-text for each news. We explore the possibilities of using Loughran-McDonald sentiment word lists [3] and Harvard IV-4 dictionary to calculate the sentiment of the news along with the usual sentiment scores given by the VADAR library. Using this idea, we generate multiple variants of the LSTM model, with features as sentiment scores obtained from various sections of the news and compare the performance with variants that don't use any sentiment scores.

Thus, there are multiple contributions from this paper: First is designing a robust and efficient news scraping and processing framework and comparing how using sentiment scores from them, as a feature in an LSTM model for stock price prediction improves the accuracy when compared to using the model without any sentiments. Second, we also analyse which section of the news article is best suitable to take sentiments from. Third, we compare the various sentiment libraries using the accuracy provided by the models with their sentiment scores as features.

The organizational structure of the paper is as follows. A few of the studies that have already been done on prediction of stock prices using LSTM and other techniques are briefly covered in Section II. The methodology and data collection process are covered in full in Section III. In Section IV, we give the outcomes of this methodology's application along with an analysis of the results. The conclusion we draw from this work and its future scope are covered in Section V.

## II. Literature review

Due to its widespread usage in business, interpersonal relationships, management, and other fields, sentiment classification and information extraction have drawn the attention of scholars from a variety of academic fields. The three basic approaches to sentiment analysis are general dictionary-specific, domain-specific [3], and machine learning techniques [4] [5]. Popular domain-specific dictionaries used for financial sentiment research include LM [6], Harvard IV, and VADER [7]. Due to their generic character, HGI, MPQA, and SenticNet are prone to incorrectly classify common financial terms. Additionally, recent studies show that employing a domain-specific lexicon rather than a broad vocabulary produces better outcomes [3].

Earlier studies primarily used newspaper [8] or Twitter [9] opinions to assess the societal backdrop of the financial market. Researchers have demonstrated that combining historical stock price models with data from popular and popular media platforms can increase the accuracy of analytical stock price prediction models [10]. The prediction of stock prices using a deep learning model like LSTM, has been found to be very accurate in [11], but it uses different forecast horizon and parameters in the model, without including sentiment scores.

Several studies, like [12], use sentiment dictionaries to extract news sentiments, forecast stock price changes, and measure market returns using models like BERT [5], LSTM. But they only focus on a single section of the article and consider whether that text is positive, neutral or negative. The proposed research differs from similar efforts in that, we have explored into considering multiple sections of the article and extracting financial sentiment polarity and intensity scores from them using three different libraries and analysing how they are improving the stock price prediction accuracy.

## III. Data and methodology

The primary purpose of this study is to determine how much better stock price predictions are when sentiment scores from company news are included as opposed to merely utilising stock prices. This is mainly achieved by preparing a *multivariate time series data*, training it using an *LSTM* model and forecasting the prices.

Following is the *architecture diagram* for our system:

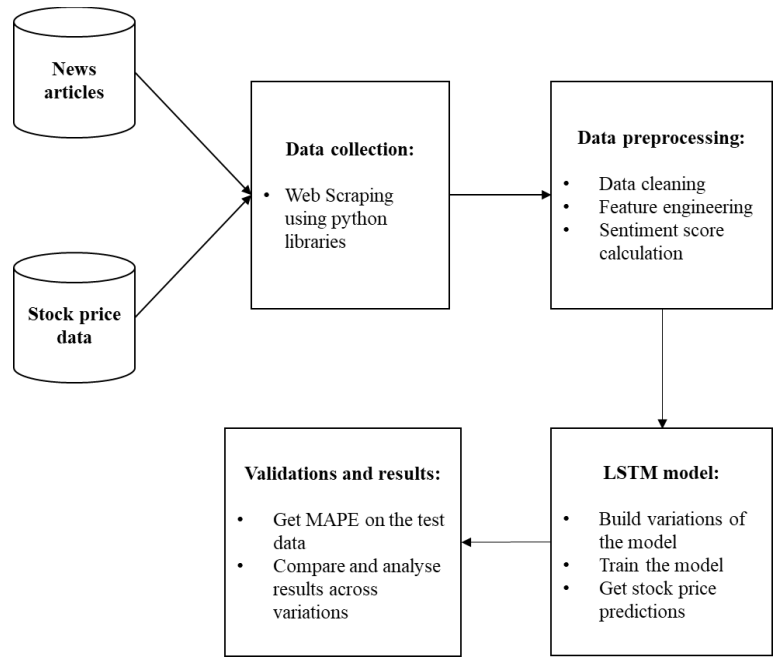

Fig 1. Architecture diagram of our system

### A. Data Collection

*Stock price* data has been collected for the latest list of *NIFTY50* stocks, from *Yahoo finance* using the *pandas_datareader* library in Python. The *timeline* we have considered for the data is from *January 2019* to *August 2022* (3.7 years). This timeline is chosen because, it's generally recommended to have at least 2 years of data to train *LSTM* models for an accurate prediction and in this case, since in 2020, there is a lot of fluctuation in the stock prices due to *COVID19*, we have considered prices for one year before also. We have a total of *904 data points*, since the stock market is not functional on weekends and hence, we are not considering the articles' sentiment scores also for these non-business dates. We also observed that, for the 904 days taken over the timeline, on an average, each stock had at least one article for 250 days, which is almost *one-fourth* of the days. Hence, we calculated the average sentiment scores for the days which have articles available for the stock and the remaining days would be filled as 0.

### B. Data Preparation

*Scraping and cleaning* all news articles for 3.7 years was a challenging task. For this, we used *Economic times* news archives and scraped the articles from its links using *BeautifulSoup* and *urllib* libraries in Python. But we could get only raw, unclean text from this. We then designed a framework to scrape out, deduplicate and clean the following *sections* separately from this: *heading, synopsis, article full-text, sector (of the company being discussed in the article), date&time of publishing of article.*

### C. Feature engineering:

To get the *sentiments* of the text, we use three *libraries*: *VADAR, Harvard IV4 and Loughran-McDonald.*

- *VADER (Valence Aware Dictionary for sEntiment Reasoning)* is a widely used library to capture the sentiments expressed in a text using valence score, describing both polarity and intensity of the emotions expressed. Below is the formula used in calculating the *compound score s* in VADER:

$$s = \frac{v}{\sqrt{v^2 + a}}$$

 where *v = sum (valence scores of words in text)*, and *a = Normalisation constant (default value is 15)*.
 We use the *nltk* python library to get Vader sentiment compound scores.

- *Harvard-IV4* sentiment dictionary consists of a list of words, under categories like positive, negative and designed for use across a variety of contexts. Following is the formula uses to get the *polarity score s* for an article using HIV4:

$$s = \frac{pos - neg}{pos + neg}$$

 where "*pos*" is the sum count of positive words and "*neg*" is the sum count of negative words in the text.

- The *LM (Loughran and McDonald)* sentiment dictionary also contains words under categories like positive, negative, but this dictionary is more specific to the financial context, hence can be a good resource for our use case. The sentiment score formula for LM remains the same as that of HIV4.

We use the *pysentiment* python library to get sentiment polarity scores for Harvard IV4 and Loughran-McDonald. We calculate these three library scores for the heading, synopsis and full text for all the articles we have collected throughout the timeline. The sentiment scores for each article are *normalised* to the range: [-1,1].

*D. LSTM Model building and validations*

The deep neural network design known as *LSTM* is fundamentally a member of the *recurrent neural network family (RNNs)*. The presence of feedback loops distinguishes *RNNs* from other deep neural networks. *RNNs* are susceptible to *the vanishing and exploding gradient problem*, which makes the network either stop learning altogether or continue learning at a very fast rate, preventing convergence to the minimal error from ever occurring. However, because of the way *LSTM* network topologies are designed, they are perfect for simulating complex sequential data, such as texts and time series, because they never experience vanishing or growing gradient problems. These networks are made up of gates that regulate and control the flow of information via cells, and these cells store the network's historical state information. An *LSTM* network uses *forget gates, input gates, and output gates* as its three different types of gates. The *forget gates* are essential for eliminating outdated information and for remembering only the information that is pertinent at the time. The *input gates* regulate the fresh data that serves as the network's current state's input. The memory cells in the network intelligently combine the current input to the network that is received through the *input gate* with the old state information from the forget gates. The output from the network is then produced at the current slot by the *output gates*. We can think of this output as the forecasted value that the model calculated for the current slot.

We have used the *keras* library in python to implement the *LSTM* layers. The *forecast horizon* is 5 days using 10 days of *past data* and we train such batches for 3.6 years. The last 30 days of the data are set aside for *testing* and are used to compute the *MAPE score* for the methodology. We have used one *input layer*, two *hidden layers* and one *output layer* of sequential type. All layers except the output layer use the *relu activation function*. The *input layer* consists of 10 units and the two *hidden layers* consist of 5 units each. The model is fit with train data using *batch size* of 5 and 100 *epochs*. We chose these parameter values based on various trial runs on some stocks and the accuracy of the results that they gave. We initially normalise the input values using *MinmaxScaler* and transform the results back later, since this has been found to be giving good results with the LSTM models.

We run the following *6 variants* of the LSTM model:

i. Input only past Close price to predict the future Close price (*one_feature*)
ii. Input past Close price, Open price, High price, Low price, Volume to predict the future Close price (*five_feature*)
iii. Input past Close price, Open price, High price, Low price, Volume, heading sentiment score to predict the future Close price (*five_feature_senti_head*)
iv. Input past Close price, Open price, High price, Low price, Volume, synopsis sentiment score to predict the future Close price (*five_feature_senti_syn*)
v. Input past Close price, Open price, High price, Low price, Volume, article full-text sentiment score to predict the future Close price (*five_feature_senti_art*)
vi. Input past Close price, Open price, High price, Low price, Volume, heading and synopsis sentiment scores to predict the future Close price (*five_feature_senti_head_syn*)

The variants from iii. to vi. were run with each of *Vader, Harvard IV4 and LM sentiment scores*, hence we have a total of 14 runs for each of the 50 *NIFTY50* stocks.

For each of the stocks, we search in all the articles throughout the timeline, if either the ticker or company name(s) is present in the respective news section *(heading/synopsis/full-text)* and if present, we consider the corresponding sentiment score for that day. For each day, the sentiment scores obtained from all the articles related to the stock are averaged and passed as one of the features to the model.

For a given date d and a stock t, final sentiment score s for that day is given by:

$$s_t(d) = \frac{\sum_{i=1}^{t_d} f(a_i)}{t_d}$$

where $t_d$ is the number of articles in the *date d* for *stock t*, $a_i$ denotes the *ith* article of the date that talks about *stock/company t* and the *function f* returns the sentiment score for a given article section.

Moreover, we reshape the Input as a *3D vector with dimensions (no. of samples, time steps (no. of past days considered), no. of features)* according to the variant that we are running. Output data will just be a *1D vector* with dimension *(no. of samples)* since we're only predicting the *Close price*. After fitting the train data from this in the LSTM model, we do prediction on the test data, based on which we calculate the *MAPE (Mean absolute percentage error)*. The formula for *MAPE* is:

$$MAPE = \frac{100\%}{n} \sum_{t=1}^{n} \frac{|A_t - F_t|}{|A_t|}$$

where $F_t$ is the predicted value and $A_t$ represents the actual value. The real value of $A_t$ is divided by their difference. For each predicted time point, the absolute value of this ratio is added, then divided by the n-th testing point. Therefore, the lower the MAPE score, the more precise our forecasts are. Because it is unaffected by the scale of the price of each stock and would standardise the comparison in percentage terms between the several variants being employed, we recommend this metric instead of others like RMSE.

The methodology that we've explained was implemented in *Python* notebooks and run in the *Google Collab GPU environment*. The run for each variant took around *4 minutes* to execute and give results.

## IV. RESULTS

Like we mentioned in *Section III*, the evaluation metric we have used to test our methodology is *MAPE*. Hence for each of the *NIFTY50* stocks, we ran the 14 variants and compared the *MAPE* among them. The main result we could obtain from that is that, *out of 50 stocks, for 40 of the stocks, variant with some sentiment column/library outperformed the other variants without any sentiment scores*. Among them, 13 are from *Vader* sentiment scores, 14 from *HIV4* and 13 from *LM* library. Only for 10 of the stocks, variants without any sentiment scores outperformed the variants with sentiment scores. Following is the table with each stock, the variant with least MAPE for that stock and the corresponding MAPE:

| Stock | Best Variant | MAPE |
|---|---|---|
| ADANIPORTS | five_feature_senti_head_syn_lm | 2.0631 |
| APOLLOHOSP | five_feature_senti_head_hiv4 | 2.4128 |
| ASIANPAINT | five_feature_senti_art_lm | 2.7351 |
| AXISBANK | one_feature | 1.9341 |
| BAJAJ-AUTO | five_feature_senti_art_lm | 2.1298 |
| BAJAJFINSV | five_feature | 3.2693 |
| BAJFINANCE | five_feature | 2.6468 |
| BHARTIARTL | five_feature_senti_art_lm | 1.8227 |
| BPCL | five_feature_senti_head_vader | 2.4137 |
| BRITANNIA | five_feature_senti_art_vader | 1.2621 |
| CIPLA | one_feature | 1.2931 |
| COALINDIA | five_feature_senti_syn_lm | 1.7076 |
| DIVISLAB | five_feature_senti_head_hiv4 | 1.9565 |
| DRREDDY | five_feature_senti_art_vader | 2.0735 |
| EICHERMOT | five_feature_senti_syn_hiv4 | 4.8554 |
| GRASIM | five_feature | 1.7299 |
| HCLTECH | five_feature_senti_art_vader | 2.1409 |
| HDFC | five_feature_senti_syn_lm | 1.6474 |
| HDFCBANK | five_feature_senti_head_vader | 1.2997 |
| HDFCLIFE | five_feature_senti_art_hiv4 | 2.0672 |
| HEROMOTOCO | five_feature_senti_art_hiv4 | 1.6457 |
| HINDALCO | five_feature_senti_syn_vader | 3.1011 |
| HINDUNILVR | five_feature_senti_art_lm | 1.5017 |
| ICICIBANK | five_feature_senti_head_syn_vader | 2.3382 |
| INDUSINDBK | five_feature_senti_head_lm | 3.7842 |
| INFY | five_feature_senti_head_vader | 2.5125 |
| ITC | five_feature_senti_head_hiv4 | 0.9753 |
| JSWSTEEL | five_feature_senti_head_syn_vader | 1.7671 |
| KOTAKBANK | five_feature_senti_head_syn_vader | 1.7278 |
| LT | one_feature | 2.4621 |
| M&M | five_feature_senti_head_syn_hiv4 | 2.4302 |
| MARUTI | five_feature_senti_head_syn_lm | 3.0227 |
| NESTLEIND | five_feature_senti_head_syn_hiv4 | 1.3042 |
| NTPC | five_feature_senti_head_hiv4 | 1.9362 |
| ONGC | five_feature_senti_head_syn_lm | 2.4579 |
| POWERGRID | five_feature_senti_art_lm | 1.5823 |
| RELIANCE | five_feature | 1.7804 |
| SBILIFE | five_feature_senti_syn_hiv4 | 2.4538 |
| SBIN | five_feature_senti_head_syn_vader | 2.0399 |
| SHREECEM | five_feature_senti_syn_hiv4 | 1.1873 |
| SUNPHARMA | five_feature_senti_head_syn_lm | 1.5617 |
| TATACONSUM | five_feature_senti_head_hiv4 | 1.7014 |
| TATAMOTORS | one_feature | 3.5610 |
| TATASTEEL | five_feature_senti_art_vader | 2.7898 |
| TCS | five_feature_senti_art_lm | 2.3405 |
| TECHM | one_feature | 1.8832 |
| TITAN | five_feature_senti_syn_hiv4 | 1.5424 |
| ULTRACEMCO | five_feature_senti_head_syn_vader | 1.6861 |

| UPL | five_feature_senti_head_syn_hiv4 | 2.4584 |
| WIPRO | five_feature | 2.1662 |

Regarding the *article sections*, we found that variant with full-text sentiment score as a feature, outperformed other variants for 12 of the stocks (*Stocks winner count*). Similarly, the variant that uses both Heading and Synopsis sentiment scores as features, outperformed other variants for 12 of the stocks. The variant using only Heading sentiment score as a feature, outperformed other variants for 9 of the stocks, while the variant using only Synopsis sentiment score as a feature outperformed other variants for 7 of the stocks. Following is the table corresponding to this result:

TABLE II.

| Article section | Stocks winner count | Average MAPE |
|---|---|---|
| Full text | 12 | 2.007601 |
| Heading & Synopsis | 12 | 2.071441 |
| Heading | 9 | 2.110244 |
| Synopsis | 7 | 2.356436 |

Comparing only the variants with sentiment scores, *across the three libraries*, we could see that, for 14 of the stocks, variants with *HIV4* sentiment score columns outperformed other variants, *Vader* for 13 of the stocks and *LM* for 13 of the stocks. Following table explains the breakdown of this result:

TABLE III.

| Library | Stocks winner count | Average MAPE |
|---|---|---|
| HIV4 | 14 | 2.066197 |
| VADER | 13 | 2.088643 |
| L&M | 13 | 2.181281 |

## V. CONCLUSION

In this paper, we have mainly explored how the addition of *sentiment scores* of stock news as a feature in an *LSTM* model, can help improve the prediction of its prices and we have clearly seen that, for at least eighty percent of the NIFTY50 stocks, these models outperform the models without any sentiment scores. We have also compared the usage of various *libraries* to get the sentiment scores, *Vader, Harvard IV4 and Lauren & Mcdonald* and found *Harvard IV4* to be the best performing and *Vader* coming closer to it. *L&M* performed particularly well with full-text, can be because it is a library more specific to financial context. Among the *sections of the article*, we found that sentiment scores from *full-text* to give the best predictions, while the variant with the combination of heading & synopsis scores coming closer to it, when compared to the variant with heading alone or synopsis alone and it can be due to the lack of news context in *heading or synopsis* alone.

As a *future scope*, we are planning to extend the experiments by adding sentiment scores to other *deep learning* models like *GRU* and other Time series methods like *ARIMA*, etc. and compare the performances. Moreover, we can try to *extrapolate* the prices and sentiment scores for the missing data on non-business days in an optimised way.